# Under-liquid Self-Assembly of Submerged Buoyant Polymer Particles[1]

Victor Multanen[a], Roman Pogreb[b], Yelena Bormashenko[a,b], Evgeny Shulzinger[b], Gene Whyman[b], Mark Frenkel[b], Edward Bormashenko[a,b]

[a]Ariel University, Chemical Engineering and Biotechnology Department, 40700, P.O.B. 3, Ariel, Israel

[b]Ariel University, Natural Science Faculty, Physics Department, 40700, P.O.B. 3, Ariel, Israel

**ABSTRACT**

The self-assembly of submerged cold-plasma-treated polyethylene beads is reported. The plasma-treated immersed millimetrically-sized polyethylene beads formed well-ordered 2D quasi-crystalline structures. The submerged floating of "light" polyethylene beads is possible due to the energy gain achieved by the wetting of the high-energy plasma-treated polymer surface prevailing over the energy loss due to the upward climb of the liquid over the beads. The capillary "immersion" attraction force is responsible for the observed self-assembly. The observed 2D quasi-crystalline structures demonstrate "dislocations" and "point defects". Mechanical vibration of self-assembled rafts built of polyethylene beads leads to the healing of "point defects". The immersion capillary lateral force governs the self-assembly, whereas the elastic force is responsible for the repulsion of polymer beads.

## INTRODUCTION

The processes of self-assembly of colloidal particles play a crucial role in the manufacturing of novel meta-materials, such as photonic crystals.[1-3] Self-assembly is a process in which components, either separated or linked, spontaneously form ordered aggregates. Self-assembly can occur with components having sizes from the molecular to the macroscopic, provided that appropriate conditions are met.[4-5]

The interest in colloidal systems was essentially strengthened when inexpensive manufacturing of monodispersed colloidal spheres became possible, leading to a diversity of promising applications.[6] It was demonstrated that these spheres form

---

perfect long-range-ordered structures.[7-8] The nature of these forces may be very different, including capillary,[9-12] electrostatic,[13-16] and electrostatic double-layer interactions.[17] In our paper we present the possibility to build long-range ordered structures based on particles lighter than a liquid and totally submerged in this liquid. Thus, 2D crystals built of polymer balls are produced, which are similar to the soaped water bubbles-based crystal structures reported by Bragg et al.[18]

## 2. EXPERIMENTAL SECTION

Spherical polyethylene beads (PB) with a radius of 0.46±0.05mm and a density $\rho$ of 960 kg/m$^3$, supplied by Cospheric LLC (USA), were exposed to a cold radiofrequency (RF) plasma treatment according to the following protocol. A RF plasma unit EQ-PDC-326 (manufactured by MTI Co, USA) was equipped with a dry vacuum pump (Edwards nXDS6i, Czech Republic) operating at a power of 18 W and a pressure of 1Torr. The Petri dish with PE beads dispersed in one layer was exposed to the plasma for 2 minutes.

The PBs were immersed into 10 vol.% water/ethanol solutions; the density of the solutions was 987 kg/m$^3$.

The "as placed" contact angles[19] and the contact angle hysteresis[19] were established with the needle-syringe method by Ramé-Hart goniometer (Model 500, USA) equipped with Auto Dispensing System 100-22-100 Ramé-Hart (USA). Droplets of 10 vol.% aqueous ethanol solution of a volume of 5 µL were placed on the extruded polyethylene films. Maximal, minimal and "as placed" contact angles, which are supposed to be close to the equilibrium (Young) angle, were measured.

The process of formation of 2D crystal structures was visualized with a Sony Exmor BSI camera, Ramé-Hart goniometer (Model 500, USA) and a SWIFT M4000-D Bifocal Microscope equipped with the digital camera UI-1485LE-C (Germany). The experiments were carried out under ambient conditions.

The vibration of floating self-assembled rafts built of PBs was carried out with the experimental device depicted in **Figure 1.** It consists of the vibrational generator Frederiksen 2185.00 (Denmark) driven by the function generator FG1617 Sampo Corp. (Taiwan); the vibration of rafts was recorded by the Sony Exmor BSI camera.

The software allowing calculation of the total surface occupied by defects from the images was developed. The software exploited the contrast between points, occupied by beads and those corresponding to water clearings. It should be emphasized that the total surface occupied by water clearings is larger than the surface occupied by defects, due to the fact that it comprises the surfaces of clearings, separating the beads, inevitable under the closed-packed ordering. The total surface occupied by defects $S$ was normalized by the entire surface $S_e$ occupied by both PBs and clearings.

**RESULTS AND DISCUSSION**

The non-treated PBs, which were lighter than the aqueous ethanol solution (the densities are supplied in the Experimental Section), floated partly protruding out of the liquid, as shown in **Figure 2A** (the actual contact angle $\theta$, shown in Figure 2A depends on the contact hysteresis to be discussed below), whereas the cold plasma-treated PBs floated being fully immersed, as shown in **Figure 2B**. The change in the light reflection manifesting itself in the disappearance of flecks is shown in **Figure 3**, evidences a change in a wetting regime. The height $h$ of the beads emersion was established with the goniometer as *ca.* 35±5 µm (see **Figure 2C**). Well-ordered 2D crystalline structures were recognized for both floating (depicted in **Figure 3A**), and submerged (shown in **Figure 3B and Figure 4**) regimes of floating.

Thus, we observed the submerged floating of beads which are lighter than a supporting liquid. The physical reasoning for this paradoxical kind of floating was supplied in Ref. 20, in which submerged floating of plasma-treated light polymer films was reported and discussed. It was explained as follows: the energy gain achieved by the wetting of the high-energy plasma-treated polymer surface prevailed over the energy loss due to the upward climb of the liquid over the polymer film. In our case we observed the additional manifestation of the same phenomenon occurring for spherically shaped light bodies. The submerged floating of light balls becomes possible when the condition:

$$\gamma_{SA}^{PT} > \gamma_{SL}^{PT} \tag{1}$$

takes place, where $\gamma_{SA}^{PT}$ and $\gamma_{SL}^{PT}$ are the specific surface energies of the plasma-treated polyethylene solid/air and solid/liquid interfaces, respectively. The validity of this inequality is well-expected. Indeed, it is well-known that the cold plasma treatment essentially increases the specific surface energy of polymer/air interfaces, and converts inherently hydrophobic polymers into hydrophilic ones.[21-22] Whereas, the specific surface energy of the plasma-treated polymer/solution interface should necessarily be low, due to the pronounced electrical charging of polymers by plasmas,[23-24] because it is energetically favorable for highly-polar water and ethanol molecules to be located in the vicinity of the charged PB surfaces, thereby decreasing the polymer/liquid specific surface energy.

The submerged plasma-treated PBs formed well-ordered long-range rafts depicted in **Figure 4**. Optical microscopy evidenced the direct contact between beads for plasma-treated PBs and clearances appearing for non-treated PB, as depicted in **Figures 5A-B**. Thus, the repulsive force in the case of submerged floating is of the elastic origin.

Consider now the nature of physical forces responsible for the formation of 2D crystalline structures, shown in **Figures 3-4**. It should be stressed that these mechanisms are different for non-treated and plasma treated PBs. Note that polyethylene is known as a strongly hydrophobic polymer, characterized by the high water Young contact angle. However, for the 10 µl droplet of a 10 vol.% aqueous ethanol solution placed on a polyethylene substrate; the advancing, $\theta_{adv}$, receding, $\theta_{rec}$ and "as placed", $\theta$, contact angles[19] are lower and were established as:

$$\theta_{adv} = 97.6° \pm 1.0°; \quad \theta_{rec} = 51.7° \pm 1.0°; \quad \theta = 87.3° \pm 1.0°.$$

And after plasma treatment these angles were:

$$\theta^P_{adv} = 47.8° \pm 1.0°; \quad \theta^P_{rec} = 22.5° \pm 1.0°; \quad \theta^P = 33.0° \pm 0.6°.$$

The high contact hysteresis governing floating of non-treated PBs, depicted in Figure 2A, is noteworthy. It is reasonable to relate the ordering of the non-treated polyethylene beads to the flotation forces, whereas attraction of submerged particles is governed by the so-called immersion capillary forces.[9-12] Flotation forces are lateral forces between particles not fully immersed into liquid, while immersion forces are lateral forces acting between particles which have no solid/air interface, as in **Figure 6**.

These two kinds of forces exhibit similar dependence on the inter-particle separation but very different dependencies on the particles' radius and the surface tension of the liquid.[9-12] The immersion forces appear not only when particles are supported by solid substrates, but also in thin fluid films.[9-12] In our experimental situation the plasma-treated PBs float beneath the liquid/vapor interface and the Archimedes buoyancy acts on them from below, instead of the supporting solid substrate. This kind of floating distorts the liquid/vapor interface, as shown in **Figure 6**; this distortion in turn gives rise to the capillary attraction force (the immersion force) acting between two soaped water bubbles of a radius $R$, separated by a distance $L$ (see **Figure 6**). The attraction force acting between fully immersed PBs has the same origin.

The self-assembly of plasma-treated PBs resembles the fascinating ordering of air bubbles floating on the surface of a soap solution in the dynamic model of a crystal structure proposed by Bragg *et al.*[18,25-27] The "point defects" and "dislocations" inherent for crystalline structures were also observed, as depicted in **Figure 3B** (see also Ref. 18).

The theory of capillary interaction developed by Kralchevsky *et al.* (Refs. 9-12) predicts the following expression for the lateral capillary force acting between two particles of radii $R_1$ and $R_2$ separated by the center-to-center distance $L$ (Refs. 9-12):

$$F = 2\pi\gamma Q_1 Q_2 q K_1(qL)\left(1+O(q^2 R_k^2)\right), \quad r_k \ll L \tag{2}$$

where $Q_k$ is the so-called "capillary charge" introduced by Kralchevsky *et al*. According to Refs. 9-12:

$$Q_k = r_k \sin\varphi_k; \quad (k=1,2) \tag{3}$$

where $r_k$ are the radii of the contact (triple) line of the particles, $\varphi_k$ is the meniscus slope angle shown in **Figure 2A**. The parameter $q$ in Eq.2 is defined according to:

$$q^2 = l_{ca}^{-2} = \frac{\rho g}{\gamma} \quad \text{(in thick films)} \tag{4}$$

$$q^2 = \frac{\rho g - \Pi'}{\gamma} = l_{ca}^{-2} - \frac{\Pi'}{\gamma} \quad \text{(in thin films)}, \tag{5}$$

where $l_{ca}$ is the capillary length, $\rho$ is the density of a fluid and $\Pi'$ is the derivative of the disjoining pressure with respect to the film thickness;[28-32] $K_1(x)$ is the modified Bessel function of the second kind of the first order. The thickness of liquid film $e$ (depicted in **Figure 6**) for the submerged regime of floating remains unknown, so the true role of the disjoining pressure and its influence on the attraction force could not be established.

The asymptotic behavior of Eq. 2 for $qL \ll 1$ is given by the expression:

$$F = 2\pi\gamma \frac{Q_1 Q_2}{L}; \quad r_k \ll L \ll q^{-1}, \tag{6}$$

which resembles a two-dimensional Coulomb law. Note that the immersion forces acting between particles when gravity is of less importance, and flotation forces driven by gravity and buoyancy, exhibit the same functional dependence on the inter-particle distance (see Eqs. 2, 6). On the other hand, their different physical origins result in

different magnitudes of capillary charges inherent to these two kinds of capillary forces, namely[9-12]:

$$F \approx \frac{R^6}{\gamma} K_1(qL) \text{ for the flotation force} \quad (7a)$$

$$F \approx \gamma R^2 K_1(qL) \text{ for the immersion force.} \quad (7b)$$

It is noteworthy that the approach developed in Refs.10-12 is not completely suited for our experimental situation, due to the fact that Eq. 2 implies the existence of the triple line, whereas in our case beads are totally covered by liquid, as depicted in Figure 2B. The dependence of the immersion force on the surface tension and the bead radius, inherent for our experiments, may be understood from the following reasoning: the potential energy of capillary interaction may be presented as $\Delta\rho g V h(L)$ where $\Delta\rho$ is the difference between the densities of immersed solid and liquid, $V$ is the volume of the bead, and $h(L)$ is the profile of the meniscus. The vertical force is $-\Delta\rho g V h'(L)$ where $h'(L)$ is the derivative over $L$. The equilibrium in the vertical direction demands compensation of the difference between buoyancy and gravity forces by the Laplace pressure, exerted from above on the fully immersed bead, supplied by: $-\Delta\rho g V = \kappa\gamma\widetilde{S}$, with $\kappa$ being the average curvature of the meniscus above the emerged part of the bead and $\widetilde{S}$ being the effective area of this part of the surface. Finally, we obtained for the capillary attraction force the dependence qualitatively resembling that in Eq. (7b):

$$F \approx \gamma\kappa\widetilde{S}h'(L) \sim \gamma\kappa R^2 h'(L) \quad (7c)$$

where the natural proportionality of $S$ to $R^2$ is supposed. Comparing Eq.7a to Eqs.7b-c we recognize that the flotation force decreases, while the immersion force increases, when the surface tension $\gamma$ increases. In addition, the flotation force decreases much more strongly than the immersion force with the decrease of the particle radius $R$.

In order to check the dependence of the self-assembly on the surface tension, we performed the experiments with non-treated and plasma treated PBs placed on soaped water. Indeed, the floating of non-treated PBs gave rise to less "defected" 2D structures on soaped aqueous alcohol solution, demonstrating the surface density of point defects smaller than that of rafts obtained on pure aqueous alcohol solution. This corresponds to the increase in the attraction force, described by Eq.7a. The plasma-treated PBs did not formed self-assembled 2D rafts on soaped water, according to predictions of Eqs. 7b, 7c.

These observations evidence indirectly that the self-assembly of non-treated PBs is governed by the flotation force (see Eq. 7a), whereas the attraction of plasma-treated PB takes place according to the immersion mechanism described by Eqs. 7b, 7c.

The difference in the behavior of plasma-treated and untreated PB was also manifested in the following experiment. A mixture of beads from the two above kinds was put in 10% water-alcohol solution. Each kind of beads formed their own cluster. The clusters of untreated beads were partially immersed (floating), while those of plasma-treated beads were fully immersed and located under the clusters of the former type forming some quasi-3D structure (see **Figure 7**). The typical cluster contained 2-3 layers of beads.

After *ca.* 20 min, the clusters of different types separated slowly. Such behavior may be understood if one takes into account that plasma-treated beads possess a "hydrophilic" surface while the surface of untreated beads is "hydrophobic" (different signs of angles $\varphi$ in (3), (6)). Accordingly, the beads of each type attract one another, but beads of different types repel. After the separation, the clusters remained stable for at least 20 hours.

As it was reported in Ref. 18, the crystalline structures formed under the capillary attraction of soap bubbles demonstrate "dislocations" and "point defects". We already considered "pseudo-dislocations" and "point defects" in the self-assembled rafts built of submerged PBs, as illustrated in **Figure 3B**. Such irregularities in metals are usually removed by slow heating called annealing, which may be modeled for PB clusters by vibration. For this purpose, the experimental set shown in **Figure 1** was designed and built. The frequency of vertical vibrations *f* was gradually increased, and the threshold value of frequency at which the "point defects" started to disappear was fixed. As is seen from **Figure 8**, after the vibration treatment with a frequency of *f*=49 Hz and amplitude of *A*=0.75 mm for 3 min, 10% of defects for untreated PBs and 30-40% of defects for plasma-treated PBs disappeared. The inspection of **Figure 8B** shows that clusters which do not contain point defects grow in size and become denser. The elaborated software (see the Experimental Section) enabled the qualitative analysis of images, and represented the kinetics of defects healing in the course of vibration (see **Figure 9**). As it is seen from **Figure 9,** the part of water clearings $S/S_e$ decreased more than two times at the end of vibration; however, it did not attain the value inherent for the close-packed 2D ordering of perfect spheres, shown in **Figure 9** with the dashed line.

Consider the quantitative information which may be extracted from the healing of defective 2D pseudo-crystalline structures by mechanical vibration. It is reasonable to suggest that healing occurs when the specific kinetic energy supplied to the 2D lattice $E_k \cong \frac{mv^2}{2}\frac{1}{\pi R^2} = \rho\frac{4}{3}\pi R^3 \frac{A^2(2\pi f)^2}{2\pi R^2} = \frac{8\pi^2}{3}\rho A^2 R f^2$ ($f$ and $A$ are the frequency and the amplitude of vibrations, correspondingly, and $m$ is the mass of PB) becomes comparable to the "energy of healing" $E_h$. For untreated PBs the energy of healing may be estimated as the energy of the adhesion of a solid particle to the liquid surface: $E_h \cong \gamma_{sol}(1+\cos\theta)^2$.[33-35] It is instructive to introduce the dimensionless parameter $\varsigma = \frac{E_k}{E_h}$, describing the ability of vibration to heal "point defects". The expression for $\varsigma$ is given by:

$$\varsigma = \frac{E_k}{E_h} \cong \frac{8\pi^2}{3}\frac{\rho A^2 R f^2}{\gamma(1+\cos\theta)^2}. \tag{8}$$

Substituting in Eq. 8 the above-mentioned experimental parameters, and $\theta = 87.3°$, $\gamma_{sol} = 49 \text{mJ/m}^2$ (Ref.36) yields an estimation $\varsigma \cong 0.3$, which is the reasonable value predicting the ability of mechanical vibrations to heal "point defects". Equation 8 predicts that the value of $\varsigma$ will increase with an increase in $R$ and decrease with an increase in $\gamma$. We plan to check these predictions in our future experiments. The energy of healing for plasma-treated PBs $E_h$ is evidently smaller (the percentage of healed defects for plasma-treated PBs was three times larger than that of the untreated, at the same parameters of vibration); however, its accurate estimation calls for future investigations.

Note that the electrostatic attraction between submerged PBs is negligible, due to the spherically symmetrical distribution of surface charges necessarily occurring for totally immersed particles, as shown in **Figure 2B**. In contrast, for the over-liquid floating it may be essential.[13-17]

It is noteworthy that Vella *et al.* calculated elastic properties (Young's modulus and Poisson ratio) of closed-packed, floating rafts[37], similar to those reported in the manuscript. It looks worthwhile to establish also the elastic properties of immersed rafts, such as depicted in Figs. 3B-4. We plan to study these properties in our current research.

**CONCLUSIONS**

We observed the under-liquid floating of cold plasma-treated light polyethylene beads immersed in aqueous ethanol solutions. The cold plasma treatment markedly increased the specific surface/air energy of polyethylene beads and in turn markedly decreased their specific surface/liquid energy. Thus, it is energetically favorable for plasma-treated polymer beads to float in a submerged, immersed regime. The ensemble of plasma-treated polymer beads tends toward self-assembly, forming close-packed long-ordered 2D pseudo-crystalline structures, due to the "immersion" capillary attraction forces. The balancing repulsive force in the case of submerged floating is of an elastic origin. The observed patterns resemble 2D pseudo-crystalline structures built of soap bubbles introduced by Bragg and Nye, and these pattern also demonstrate similar "dislocations" and "point defects". The mechanisms of self-assembly of non-treated and plasma-treated polymer beads are quite different; the attraction force for the non-treated beads is of the "floating" kind, whereas for plasma-treated beads it is the "immersion" capillary force, which increases when the surface tension $\gamma$ increases, contrastingly to the flotation force. The "dislocations" and "point defects" are healed effectively under mechanical vibrations, when the energy supplied to the beads by vibrations becomes comparable to the characteristic energy of formation of the "point defects".

**ACKNOWLEDGEMENTS**

Acknowledgement is made to the donors of the Israel Ministry of Absorption for the partial support of the scientific activity of Dr. Mark Frenkel.

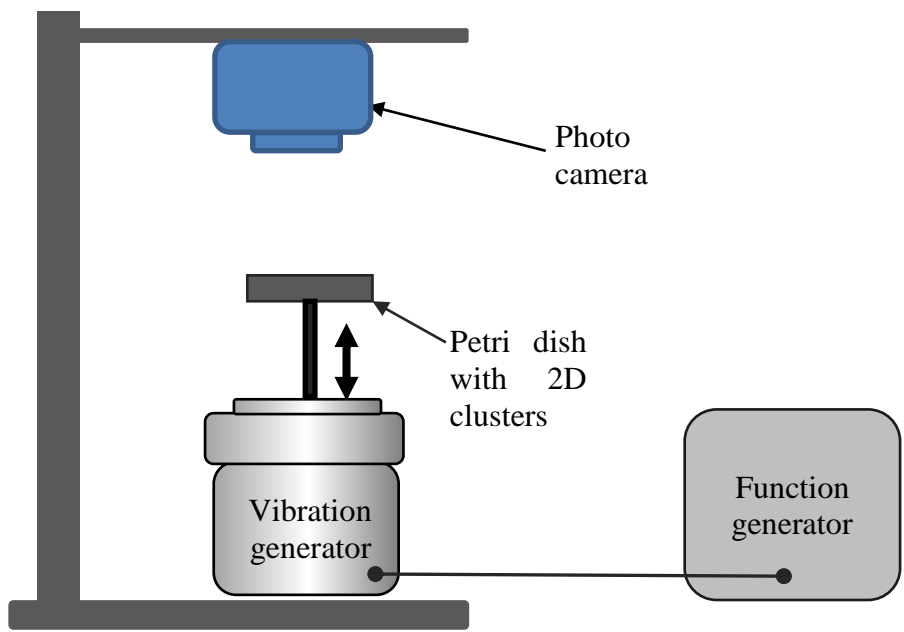

**Figure 1**. Experimental lab-made unit for vibration treatment of defects in 2D beads' clusters.

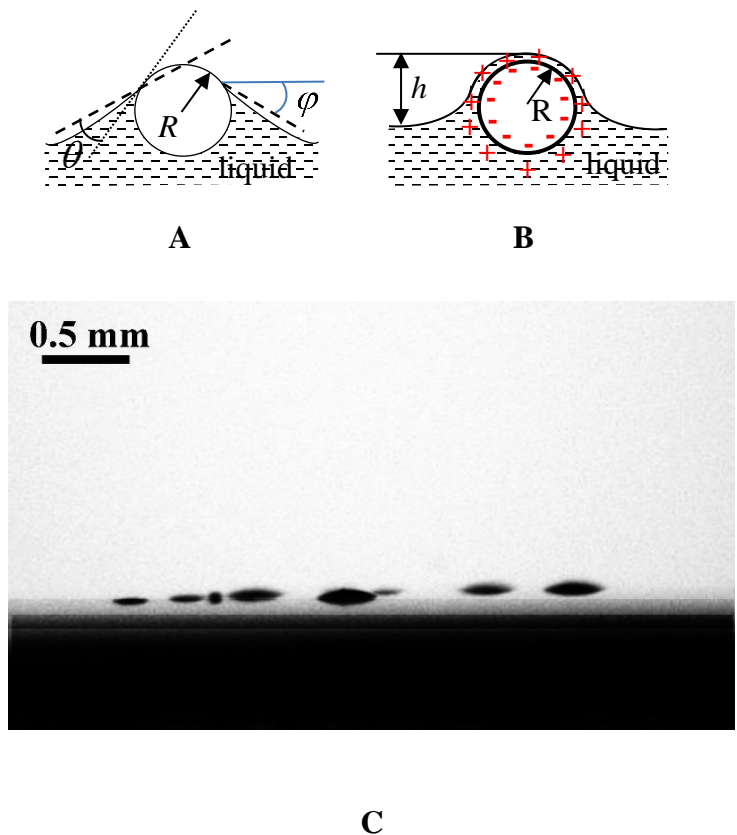

**Figure 2**. **A**. Floating of non-treated polyethylene bead. The contact angle $\theta$ is formed by the tangent to the liquid/vapor interface (the dashed line) and tangent to the PB surface (the dotted line). **B**. Submerged (under-liquid) floating of the plasma-treated polyethylene bead. **C**. Under-liquid floating of the plasma-treated polyethylene beads as seen with goniometer. The emerged parts of beads are covered with a thin liquid layer.

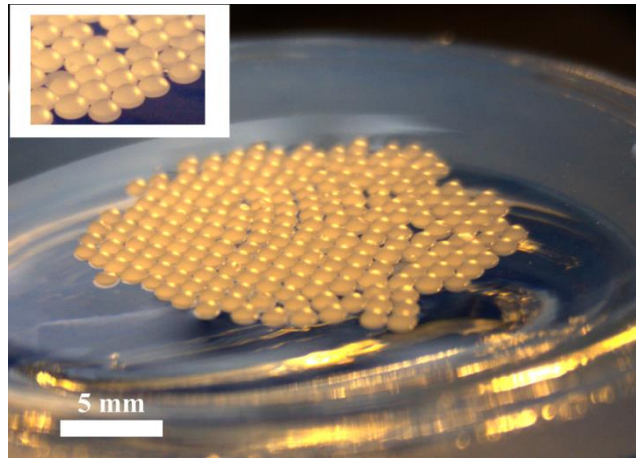

**A**

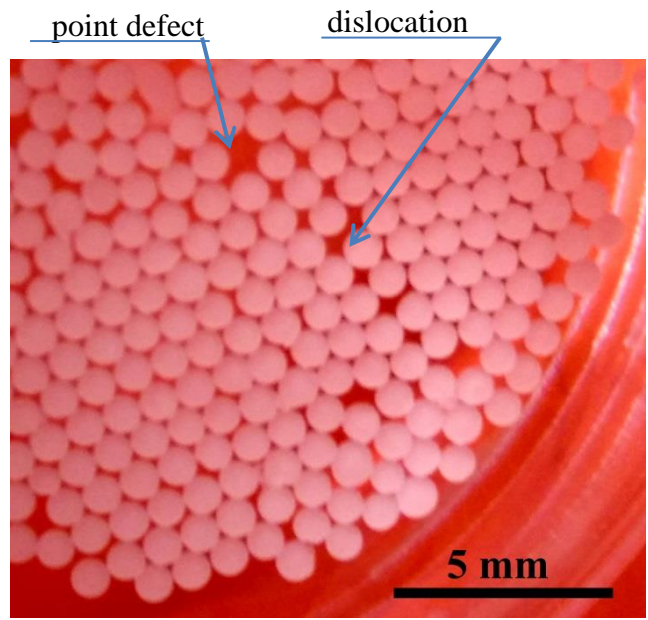

**B**

**Figure 3**. **A.** Floating of non-treated PBs. **B**. Submerged floating of plasma-treated PBs. Flecks shown in **Figure 3A** disappeared.

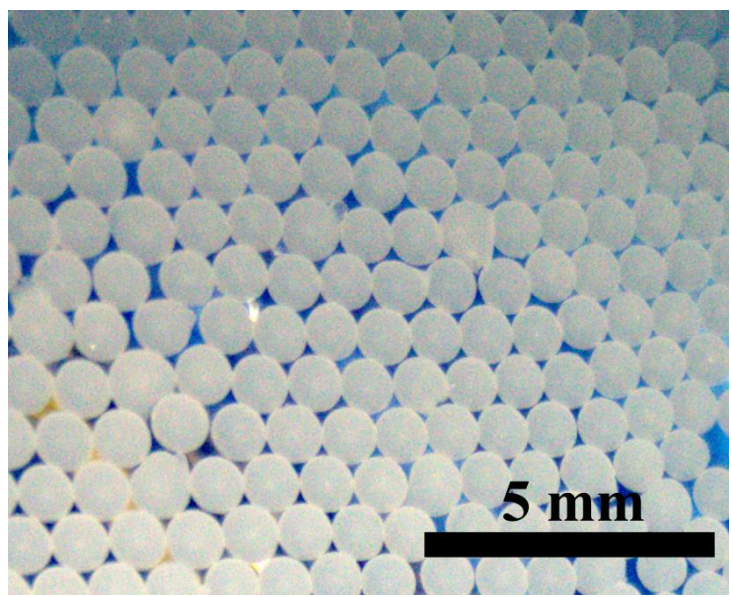

**Figure 4**. Close-packed 2D structures built of plasma-treated submerged PBs.

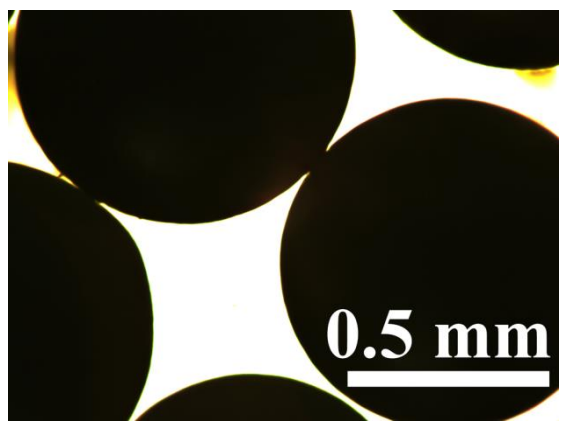

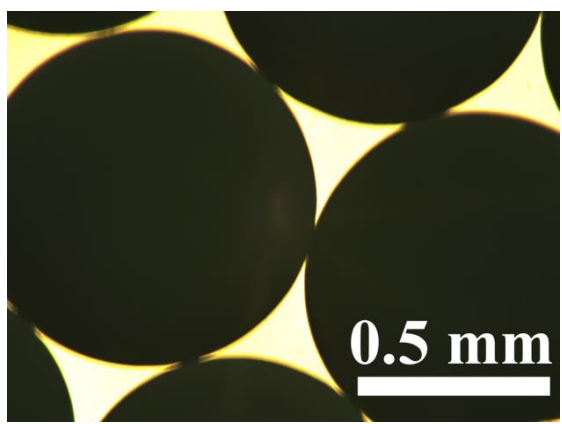

**Figure 5**. Optical microscopy images of the direct contact of floating PBs. A. Over-liquid floating of non-treated PB. B. Under-liquid floating of plasma-treated PBs.

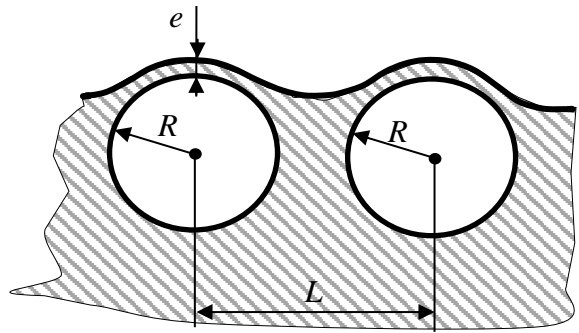

**Figure 6**. Scheme illustrating the "immersion" capillary interaction between submerged beads of the same radii $R$. The center-to-center separation of beads is $L$.

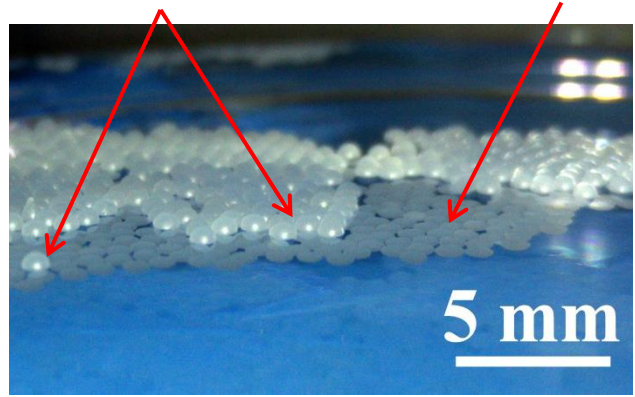

A

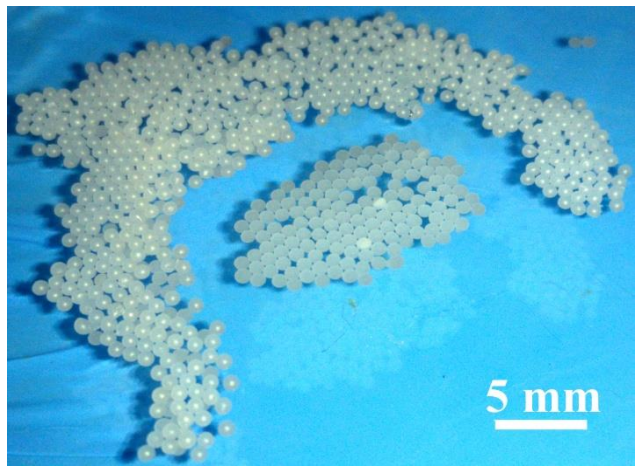

B

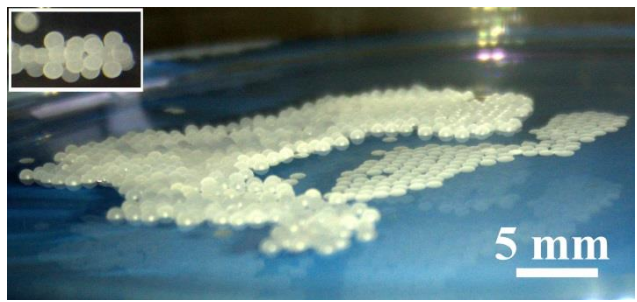

C

**Figure 7.** Plasma-treated and untreated PBs placed on the surface of aqueous ethanol solution. The plasma-treated beads look darker because they form clusters fully immersed into the solution. The untreated beads form clusters, which are not fully immersed; they look shiny and float on the surface. A. Side view. B. Top view. C. The two kinds of clusters start their separation. The initial stage is shown in insert, where the layered structure is seen.

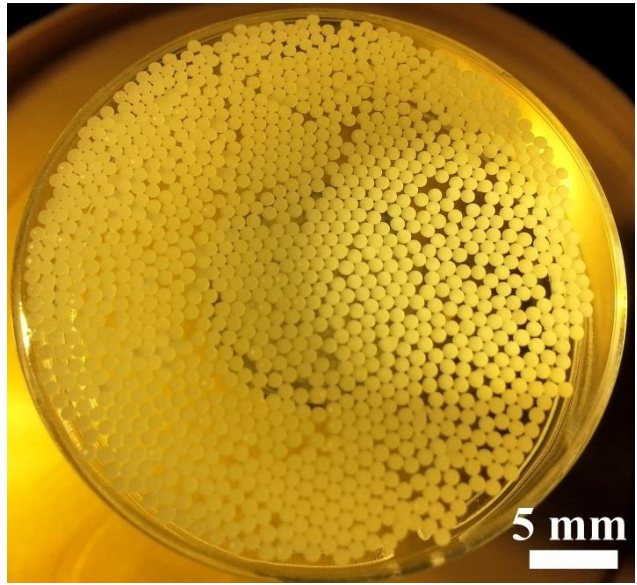

A

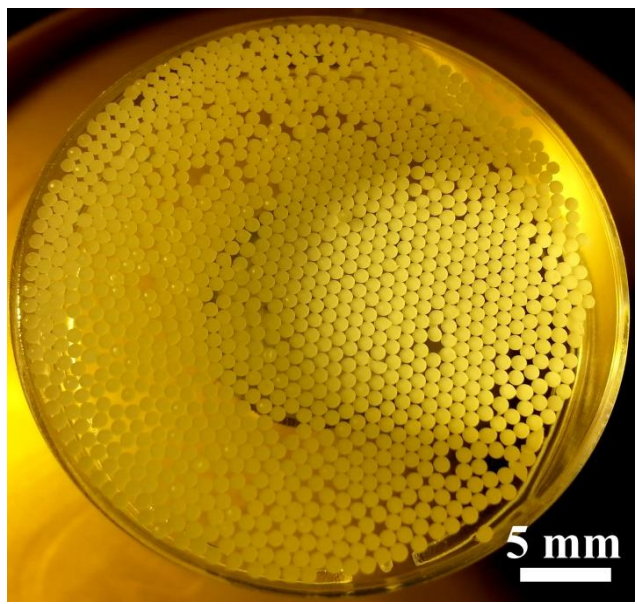

B

**Figure 8.** Removal of dislocations and point defects with vibration. A. Self-assembled two-dimensional cluster of immersed PBs. B. After three minutes of vertical vibration, the same immersed cluster contains a smaller number of defects per unit area.

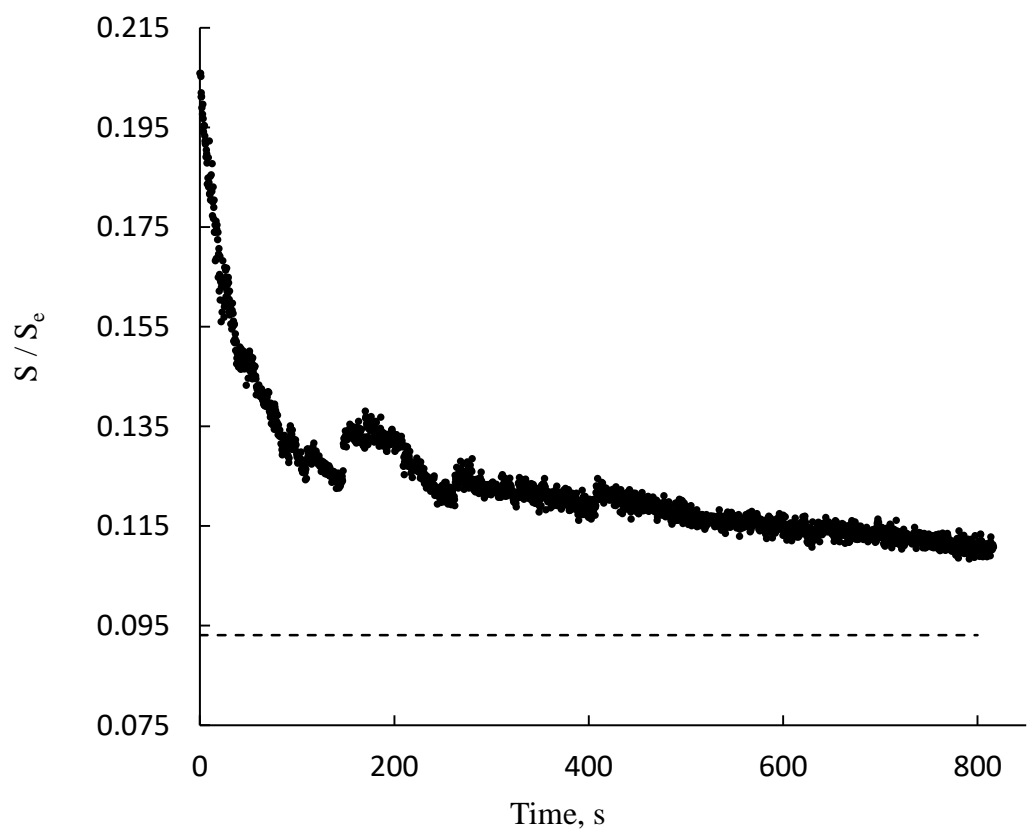

**Figure 9**. The kinetics of the change in the fraction of the free space area separating PBs S relative to the entire area Se occupied by both PBs and clearings. The dashed line supplied for guiding the eye shows the value 0.09310 corresponding to the ideally close-packed plane structure.